\documentclass[twocolumn,showpacs,preprintnumbers,amsmath,amssymb,superscriptaddress,prb,english]{revtex4-1}
\usepackage{epsfig,amsopn}
\usepackage{amssymb,epsfig}
\usepackage{epsfig}
\usepackage{graphicx}
\graphicspath{{figure/}}
\usepackage{epstopdf}
\usepackage{color}
\usepackage[caption=false]{subfig}
\usepackage{amsmath,amssymb}
\usepackage{amsthm}
\usepackage{enumerate}
\usepackage[T1]{fontenc}
\usepackage[utf8]{inputenc}
\usepackage{mathtools}   
\usepackage{empheq}

\usepackage{physics}
\newcommand{\comment}[1]{}
\newcommand\beq{\begin{equation}}
\newcommand\eeq{\end{equation}}

\begin{document}
\title{Curvature function renormalisation, topological phase transitions and multicriticality}
\author{Faruk Abdulla}
 \affiliation{Harish-Chandra Research Institute, HBNI, Chhatnag Road, Jhunsi, Allahabad 211 019, India.}
 \author{Priyanka Mohan}
 \affiliation{Department of Theoretical Physics, Tata Institute of Fundamental Research, Homi Bhabha Road, Colaba, Mumbai 400005, India.}
 \author{Sumathi Rao}
 \affiliation{Harish-Chandra Research Institute, HBNI, Chhatnag  Road, Jhunsi, Allahabad 211 019, India.}

\begin{abstract}
A recently proposed curvature renormalization group scheme for topological phase transitions defines a generic `curvature
function' as a function of the parameters of the theory and shows that topological phase transitions 
are signalled by the divergence of this function at certain parameters values, called critical points, in analogy with usual phase transitions. 
A renormalization group procedure was also introduced as a way of flowing away from the critical point towards a fixed point,
where an appropriately defined  correlation function goes to zero and  topological quantum numbers characterising the 
phase are easy to compute. In this paper, using two independent models - a model in the AIII symmetry class and
a model in the BDI symmetry class -   in one dimension as examples,
we show that there are cases where the fixed point curve and the critical point  curve appear to intersect,
which turn out to be multi-critical points,
and focus on understanding its implications.

\end{abstract}

\maketitle
\section{Introduction}
The Landau order parameter paradigm \cite{landau1937, miransky} describes continuous phase transitions with spontaneous symmetry breaking. Systems undergoing these transitions often possess a local
 order parameter which is present only in one of the phases. Close to the phase transition point or critical point, the system exhibits self-similarity
 or scale invariance, and these transitions can be studied using Kadanoff’s scaling theory \cite{Kadanoff1966}.
    
 In the last couple of decades,   topological phase transitions have garnered a lot of  attention. These transitions fall outside 
 Landau's paradigm and cannot be described by a local order parameter.
   They are tuned by varying the coupling parameters in the model. Even though different topological models are classified based on the dimensions and the symmetries of the model,
     \cite{Haldane1988, HasanKane2010, KaneMele2005, KaneMele2005-1, Kitaev2001, ReadGreen2000, BernevigHughesZhang2006, QiZhang2011, QiHughesRaghuZhang2009} these transitions  do not involve spontaneous 
     symmetry breaking like in Landau's theory.
    The different topological phases are distinguished by a topological invariant and the discrete change of this invariant signals the transition between the phases.

In this context, a renormalization group approach has been introduced
 where a scaling procedure,  analogous 
to the Kadanoff’s scaling theory, for topological systems is derived. This is based on the following observation: The topological invariant in many cases is calculated by 
integrating a function, known as the curvature function,  over the whole Brillouin zone. This function, which diverges at the transition holds the information about the topology of the band structure. 
The concept of scaling here is to change the curvature function in such a way that the topology does not change.
 This scaling procedure is compared to that of stretching a messy string to reveal the  number and types of knots that it contains. 
The renormalization group flow is then chosen to reduce the divergence, without changing the topology,   and to finally terminate at  a fixed point. 
At this point the curvature remains unchanged under 
further renormalization. 
 Since the scaling procedure acts on a curvature function, this method has been called the curvature renormalization group (CRG) approach.

The CRG approach has been used in studying topological systems such as  the Su-Schrieffer-Heeger model \cite{SuSchriefferHeeger1979, Chen2016} and periodically driven systems \cite{MoligniniChenChitra2018, MoligniniChenChitra2019}. It has been extended to analyze topological phase transitions 
involving higher order  band crossings \cite{ChenSchnyder2019} and models with $Z_2$ topological invariants.\cite{ChenSigristSchnyder2016} A class of systems which are different from the above mentioned cases are the
ones which are not exactly solvable, such as  interacting systems. The CRG is also a useful technique for  studying  topological transitions in weakly interacting systems as shown in Ref.~\onlinecite{Chen2018}. It has also been   successfully applied to
a strongly interacting fractional Chern insulator system\cite{KourtisNeupert2017}.

 The strength of the scaling technique lies in the fact that the fixed points and critical points of the theory hold  valuable  information about the topological phase transitions (TPTs) in  the model,
 thus sparing us from scanning the whole parameter space to construct the phase diagram. 
The idea is that there may be situations, notably in higher dimensional or  interacting theories, where the direct study of the curvature (CRG)
may be more feasible than integrating the curvature function to compute the topological invariant, which requires the knowledge
of the curvature function at all points in the Brillouin zone.

In this paper, we carry forward the analysis to models,  where there exists critical lines at non high symmetry points,
besides the critical lines at  high symmetry points (HSP) in the Brillouin zone.
Although these critical lines at non high symmetric points cannot be discovered by a straightforward application of the CRG method, ( particularly
when they are not isolated), 
it is possible to analyse the system using the scaling of the curvature function and the divergence of an appropriate  correlation length.
Using this approach, we study two models in this paper. The first is the Kitaev chain with extended couplings, a model in the BDI class,
and the second is the Su-Schrieffer-Heeger model with extended couplings, a model in the AIII class. We summarise here the main
results in both the models. We find that for large fractions of the parameter space,
the unstable part of the  fixed point line overlaps with a  critical line, thereby showing that unstable fixed point lines also denote 
topological phase transitions. Hence, at the crossing point where an unstable fixed point line  meets or  intersects a critical curve, three phases coexist 
and hence  the  crossing point turns out to be a multi-critical point.

In the rest of the paper, we derive and elaborate upon these results. 
 The remaining sections are organized in the following way: In Sec.\ref{sec:curvature_rg}, 
 we describe the curvature renormalization procedure briefly. In Secs. III and IV, we apply the curvature RG procedure to the 
 extended Kitaev model  and to the extended Su-Schrieffer-Heeger model and obtain their  phase diagrams, 
 analyse their  flow equations and obtain their Wannier state correlation length. In Sec V, we discuss the common features of
 the parameter landscape which exist in both these models and analyse its implications.
 The conclusion and summary are described
 in Sec.\ref{sec:dc}.
 
\section{Curvature RG}
 \label{sec:curvature_rg}

\begin{figure}
\includegraphics[width=0.85\linewidth]{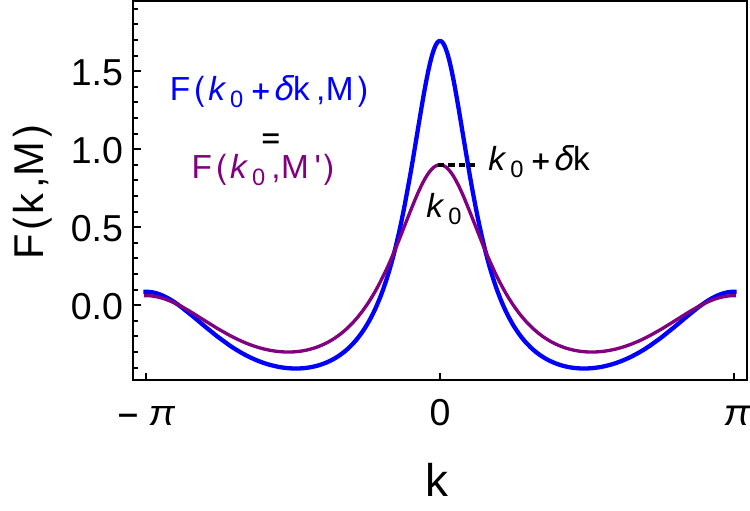}
\caption{The schematic deviation-reduction mechanism: For a given $\bf{k_0+\delta k}$ for $\bf{M}$, one has to find $\bf{M'}$ for which $F(\bf{k_0+\delta k, M})=F(\bf{k_0, M'})$ such that peak gets reduced. }
\label{fig:dev}
\end{figure}

In this section, we briefly review the curvature renormalization group method (CRG) introduced in  Refs.\onlinecite{Chen2016, ChenLegnerSigrist,  NieuwenburgSchnyderChen2018, MoligniniChenChitra2018}  and further explained in 
 Refs.\onlinecite{ChenSigristSchnyder2016,Chen2018, ChenSchnyder2019, MoligniniChenChitra2019}. 
The different topological phases in a system are distinguished by a topological invariant, which is calculated by integrating a function over the  Brillouin zone. This integrand function,
called the curvature function in the rest of the paper, can either be the Berry curvature, the Berry connection or the Pfaffian of an appropriate `sewing matrix'  as dictated by the dimensions and the
underlying symmetry class of the system\cite{Chen2016}. The topological invariant $C$ is then calculated by the following equation: 
\beq
\label{eq:invariant}
C = \int_{BZ} \frac{d^D {\bf k}}{(2\pi)^D} F({\bf k},  {\bf M} ).   
\eeq
 Here $F({\bf k},  {\bf M} )$ is the curvature function and ${\bf M}=(M_1, M_2,\cdots, M_i,\cdot) $ is the set of all the coupling parameters in the theory.

  Consider a point, $\bf{M}={\bf M}_c$ in the parameter space where the system undergoes a topological transition i.e. the topological number $C$ changes. At $\bf{M_c}$, the bulk band gap closes usually at a high symmetry point (HSP), ${\bf k}_0$, 
  in the Brillouin zone resulting in a diverging curvature function.
  For a small perturbation $\delta \bf{k}$ near ${\bf k}_0$, the CRG procedure can be summarized in the following equation:
\beq 
\label{eq:dr}
F({\bf k}_0 +\delta {\bf k}, {\bf M}) =  F({\bf k}_0, {\bf M}^{\prime}).
\eeq
Given the curvature function (LHS) at $\bf{k_0+\delta k}$ for parameters ${\bf M}$, one has to find a new  ${\bf M'}$ which makes $F({\bf k}_0, {\bf M}^{\prime})$ equal to $F({\bf k}_0 +\delta {\bf k}, {\bf M})$. As discussed  in detail in Ref.\onlinecite{Chen2016}, this 
  procedure gradually reduces the divergence of the curvature function  at $ {\bf k}_0 $
  as demonstrated in Fig.\ref{fig:dev}. This is known as the deviation-reduction mechanism.
  Under the iterative application of Eq.\ref{eq:dr},  the parameters ${\bf M}$ flow away from the critical point ${\bf M}_c$ towards a fixed  point ${\bf M}_0$.
  When the flow stops at ${\bf M}_0$, the curvature function has the form: $F({\bf k}_0 +\delta {\bf k}, {\bf M}_0) =  F({\bf k}_0, {\bf M}_0)$. 

The equation to track the flow of curvature function to its fixed point in the parameter space can be derived by expanding the RHS and LHS of Eq.\ref{eq:dr} to leading order
 in $\delta {\bf k}$ and $\delta {\bf M}=M'_i-M_{i}$. This gives,  
 \beq \label{eq:fe}
\frac{d M_i}{dl}= \frac{1}{2} \frac{{\partial^2_{k_j} F({\bf k}, M_i)}|_{{\bf k}={\bf k}_0}}{\partial_{M_i} F({\bf k}_0, M_i)}, 
\eeq 
where  $ dl= {\delta k^2_j} $,  $ \partial_{k_j}$ is the partial derivative with respect to  $ k_j$ and $ \partial_{M_i}$ is the partial derivative with respect to  $M_i$.
Note that  we have used the fact that the curvature function is an even function near a gap-closing high symmetry point, i.e. close to  $k_0$, $F({\bf k_0 +\delta k},  {\bf M} ) = F({\bf k_0 -\delta k}, {\bf M} )$.
The RHS of the above equation diverges at the critical point $M_{ic}$ and vanishes at the fixed point $M_{i0}$. Flows in the parameter space, right along the  critical line, diverge and all the nearby flows are away from the critical line of TPTs. Note that the flow Eq.3  can be solved from the knowledge of the curvature in the neighbourhood of the high symmetry point ${\bf k}_0$. Thus CRG is able to identify the topological phase boundaries
without the knowledge of the curvature function for the whole Brillouin zone.
  
 The curvature function is peaked at HSP ${\bf k_0}$ with the property  $F({\bf k_0 +\delta k},  {\bf M} ) = F({\bf k_0 -\delta k}, {\bf M} )$ , and thus expected to have a Lorentzian form (see the schematic in Fig.\ref{fig:dev}) near the critical point and therefore  can be written in the following way:
 \begin{equation}
F({\bf k}_0 + \delta {\bf k}, {\bf M}) = \frac{F({\bf k}_0, {\bf M} )}{1 + \xi _{{\bf k}_0}^2 {\delta \bf k}^2} .
\label{Eqn:oz}
  \end{equation}
where $ \xi _{{\bf k}_0}$, known as the correlation length,  is the length scale associated with the divergence  of the  curvature function at the transition\cite{ChenLegnerSigrist}. This 
can also be identified with the decay length scale  of  the Wannier state correlation function (defined as the overlap between two Wannier states which are at a distance $R$ from each other) $\lambda_R$\cite {MoligniniChenChitra2019}.  In one dimensional systems, $\lambda_R$ scales as $\lambda_R\sim e^{-R/\xi_{k_0}}$.
 
  From the divergence and the associated behavior described above, a scaling form can be attributed to the curvature function near the critical point. Therefore one can write, $|F({\bf k}_0, {\bf M})|\sim 
  |{\bf M}-{\bf M}_c|^{-\gamma}$.  Similarly using Eq.\ref{Eqn:oz}, the scaling form of the correlation length is written as $\xi_i\sim |{\bf M}-{\bf M}_c|^{-\nu_i}$ . The exponents
  $\gamma$ and $\nu_i$ are the critical exponents
  associated with this topological transition. In 1D systems they are simply related by the expression: $\gamma=\nu_i$. 
   
    So far, we have discussed a  scenario known as the peak divergence scenario. Here the curvature function develops a divergent  peak at one of the high symmetry points as we approach the transition. There is another case known as the 
    shell divergence scenario in case of higher order band crossing. Here, as we move towards the critical point in the parameter space, the curvature function peaks in the forms of  a $D-1$ 
    dimensional shell around the HSP.
    The RG formalism in this case is discussed in Refs.\onlinecite{ChenSchnyder2019,ChenSigristbook}.

As explained in this section, the CRG procedure is an iterative method to search for the trajectory in the parameter space wherein the maximum of the  curvature function reduces. In this way, we obtain
the flow equations corresponding to transition at a particular HSP in Eq.\ref{eq:fe}. Once the HSPs of a system are identified, which are usually a few, one can carry out this analysis to obtain 
the complete
flow diagram in the parameter space. As we demonstrate later in the paper, the flow diagram in the parameter space divides it into different topological regions. This removes the need to compute the topological invariant 
at each point in the parameter space. Therefore, CRG is an efficient method when the number of couplings are large. The invariant needs to be calculated only for a few points, 
which are from topologically different regions in this space.

{\section{CRG analysis of the extended Kitaev model}}
\label{Kitaevmodel}

\subsection{The model and its phase diagram}

\begin{figure}[htb]
 \includegraphics[width=0.65\linewidth, height=0.6\linewidth]{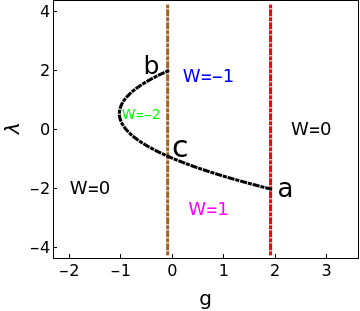}
  \caption{The phase diagram of the Kitaev model with next nearest neighbour  couplings in the $g$-$\lambda$ ($\lambda=\lambda_1/\lambda_2$) plane. The red line at $g=1.9$ corresponds to the gap closure  at $k_0=0$ and the brown line at $g=-0.1$  represents the gap closure at $k=\pi$ . The black curve represents the array of non-high-symmetry  gap closing points (for the choice $\lambda_2=0.85$ and $t=0.9$). The winding numbers ($W$) in each of these phases are also shown ($\lambda_2> 0$). The winding numbers change sign for $\lambda_2<0$. Note the existence of multi-critical points  (a) and (b) where
  phases with three different winding numbers meet at a point and the multi-critical point  (c) where four different topological phases meet.}
 \label{fig:pd}
 \end{figure}

Here, we consider the 1D Kitaev spinless p-wave superconducting chain\cite{Kitaev2001} with both nearest and next-nearest coupling terms\cite{koppchakravarthy2005,NiuChungHsuMandalRaghuChakravarthy2012,ThakurathiDiptiman2013}. The Hamiltonian is given by,  
\begin{align}
 H  & = -t_1 \sum_{\langle ij\rangle}\left( c_i^{\dagger} c_{j} + h.c \right) - t_2 \sum_{\langle\langle ij\rangle\rangle}\left( c_{i}^{\dagger}c_{j} + h.c \right)  \nonumber \\ 
 & -\lambda _{1} \sum_{\langle ij\rangle}\left( c_i^{\dagger} c_{j}^{\dagger} + h.c \right) - \lambda _2 \sum_{\langle\langle ij\rangle\rangle}\left( c_{i}^{\dagger}c_{j}^{\dagger} + h.c \right) \nonumber \\
 & + g \sum_i \left( 2 c_i^{\dagger}c_i -1\right).  
\end{align} 
where $g$ is the chemical potential, $t_{1,2}$ are the nearest(NN) and next-nearest-neighbour(NNN) hopping terms and $\lambda_{1,2}$ represent NN and NNN superconductor pairing terms respectively. We take the pairing amplitudes $\lambda_1$ and $\lambda_2$ to be real. The real pairing amplitude makes the Hamiltonian  time reversal symmetric. When the Hamiltonian is expressed in Bloch-Boguliobov-de Gennes form in momentum space, it is also  particle-hole symmetric. In the topological classification scheme, the model falls in the BDI class. This model was studied in detail in  Ref. \onlinecite{NiuChungHsuMandalRaghuChakravarthy2012} for $\lambda_1=\lambda_2$  and $t_1=t_2$ where the different topological phases  of this system were analyzed, and the topological invariant was shown to be given by the winding number $Z$.
Here  we remove the  constraint on the parameters to extend our parameter space. 
 After Fourier transformation, the  Hamiltonian can be written in the Boguliobov-de Gennes form in the basis $(c^{\dagger}_k, c_{-k})$ as
\begin{align}
 H(k)  =  & \sum_i d_i(k) \sigma_i \label{2stateham}\\
\rm{where}~~
 d_1(k) = &0 \nonumber \\
  d_2(k)= & -2\lambda _1\sin{k} - 2 \lambda _2 \sin{2k}  \nonumber \\
  d_3(k)= & 2g - 2t_1\cos{k} - 2t_2\cos{2k} .  \nonumber \\
\end{align}
 Here the  $\sigma_i$'s   are the Pauli matrices. The  energy eigenvalues are given by
\beq
\label{eq:bands}
E _k  = \pm \sqrt{ d_2(k)^2 + d_3(k)^2}.
\eeq
 The energy  gap  closes when both the squared terms in the energy vanish together for some $k$ i.e.,
  \beq \label{eq:17}
 \lambda _1 \sin{k} + \lambda _2 \sin{2k}= 0, \quad g - t_1 \cos{k} - t_2 \cos{2k}  = 0 .
 \eeq
At the two HSPs $k_0=0 $ and $k_0=\pi$, the first of the two equations above is trivially satisfied
for  all $ \lambda_{1}$ and $ \lambda_{2} $ and
the second one reduces to $g = t_2 \pm t_1 $ where the top and bottom signs are for   $k_0=0 $ and $k_0=\pi$ respectively. 
More generally, combining  the two equations gives a set of  gap closing points which are not at HSPs. 
These gap closing points are at momenta given by
\beq \label{eq:19}
k=\frac{1}{2}\arccos{\left[ \frac{t_1}{t_2}\left(\frac{\lambda}{2}+\frac{g}{t_1} \right) \right]},
\eeq 
with $\lambda=\lambda_1/\lambda_2$. In the parameter space of $(g,\lambda,t_1,t_2)$, the  non-HSP gap closing points 
are, hence,  given by the following equation:
\begin{equation}
 g=t_2\frac{\lambda^2-2}{2}-t_1\frac{\lambda}{2}~,
 \label{Eq:gl}
\end{equation}
with the constraint $-2<\lambda<2$. This is a line in  parameter space forming  the boundaries of the different phases as shown in 
Fig.\ref{fig:pd}.

 The spinless Boguliobov-de Gennes Hamiltonian $H(k)$ is time reversal (TR) invariant  $TH(k)T^{-1}=H(-k)$, with $T$ being complex conjugation\cite{NiuChungHsuMandalRaghuChakravarthy2012,ThakurathiDiptiman2013} and, in addition, due  to the  particle hole symmetry of the BdG Hamiltonian, it has the chiral symmetry  $SH(k)S^{-1}=-H(k)$\cite{ChiuTeoSchnyder2016}, where the chirality  operator is 
 given by $S=\sigma_1$. Consequently,  a  unitary transformation by
\begin{align*}
 U=\frac{1}{\sqrt{2}}\left(\begin{array}{cc}1 & -1 \\1 & 1\end{array}\right),
\end{align*}
 brings the BdG Hamiltonian to the block-off diagonal form: $ UH(k)U^{-1}= d_3(k) \sigma_1 +  d_2(k) \sigma_2$. This model exhibits non trivial topological phases\cite{NiuChungHsuMandalRaghuChakravarthy2012} distinguished by the winding number  $W$,
 \beq \label {eq:winding} 
 W =  \frac{1}{V_{BZ}}\int_{BZ} d\theta_k   =\frac{1}{2\pi} \int_{-\pi}^{\pi}  \left(\frac{d\theta _k}{dk}\right)dk ,
 \eeq
 where  $ \theta_k $ is  given by $ \tan \theta_k = d_3 (k) /d_2 (k) $. The winding numbers
 of the different phases are also  shown in Fig.\ref{fig:pd}. In particular, we note the existence of three multi-critical points marked by points `a', `b' and `c' where more than two phases with different
 winding numbers meet. The points `a' and `b' are the meeting point of three phases, while the point `c' sees the intersection point of four different topological phases. In Sec. IIIB, we will show that the 
 points `a' and `b' are interesting from the curvature RG point of view.

\begin{figure} [ht]
\includegraphics[width=0.48\linewidth, height=0.45\linewidth]{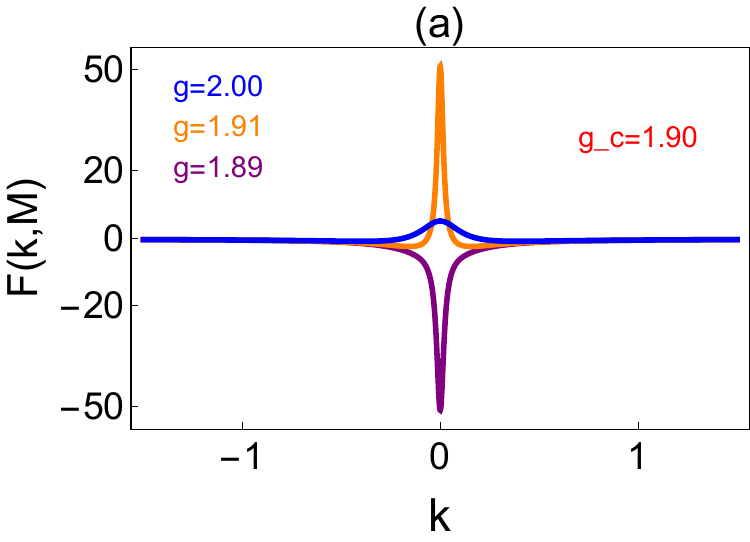}
\includegraphics[width=0.48\linewidth,  height=0.45\linewidth]{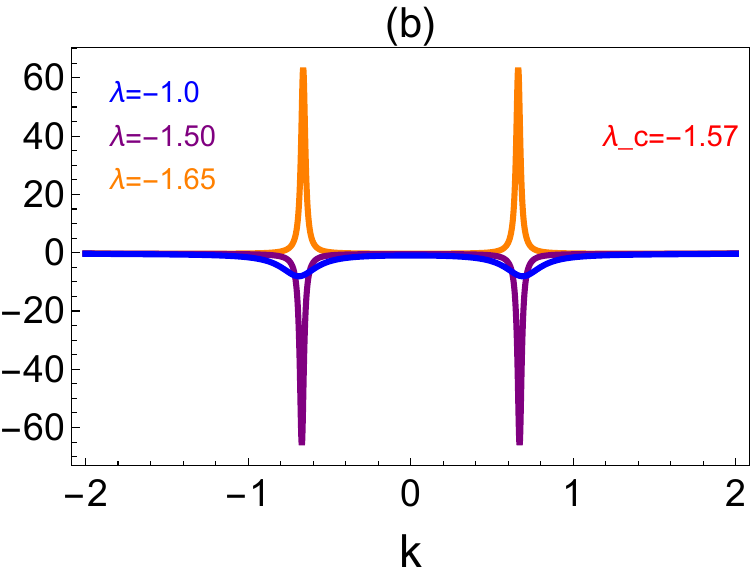}
\caption{The divergence and sign change of  the curvature function at (a) HSP critical point and at (b) non-HSP critical points for the the Kitaev model. In (a) $\lambda=-1.4$ and in (b) $g=1.0$.}
\label{Fig:curv_div}
\end{figure}

  From  Eq.\ref{eq:winding}, we read the curvature function to be, 
\begin{equation}
 \label{eq:cf}
F(k, \mathbf{M}) = \frac{d\theta_k}{dk}  = \frac{d_2(k) \partial_k d_3(k) -  d_3(k) \partial_k d_2(k) }{d_2(k)^2 + d_3(k)^2} 
\end{equation}
where,  $d_2(k) = -2 \lambda _1 \sin{k} - 2 \lambda _2 \sin{2k}$,   $d_3(k) =2g-2\cos{k} - 2t\cos{2k} $,  and  the parameters 
$\mathbf{M} =(g, t, \lambda_1, \lambda_2)$, 
 with $t_1$ being set to one and $t_2=t$. 
 We plot the curvature function in Fig.\ref{Fig:curv_div} to show its  divergence and sign flip across the  (a) HSP gap closing point and (b) non-HSP gap closing point.

  \begin{figure*}
  \centering
\subfloat[RG flows for $k_0=0$. ]
{\includegraphics[width=0.33\linewidth, height=0.3\linewidth]{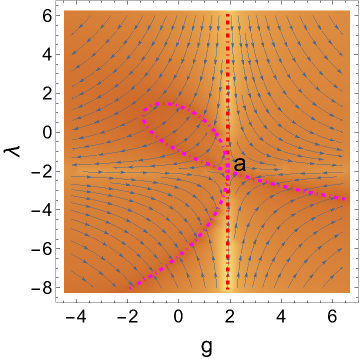}
\label{fig:flowsk0}}
\subfloat[RG flows for $k_0=\pi$. ]
{\includegraphics[width=0.33\linewidth, height=0.3\linewidth]{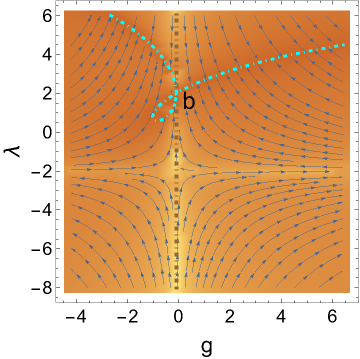}
\label{fig:flowskp}}
\subfloat[Critical and fixed lines in the model ]
{\includegraphics[width=0.29\linewidth, height=0.3\linewidth]{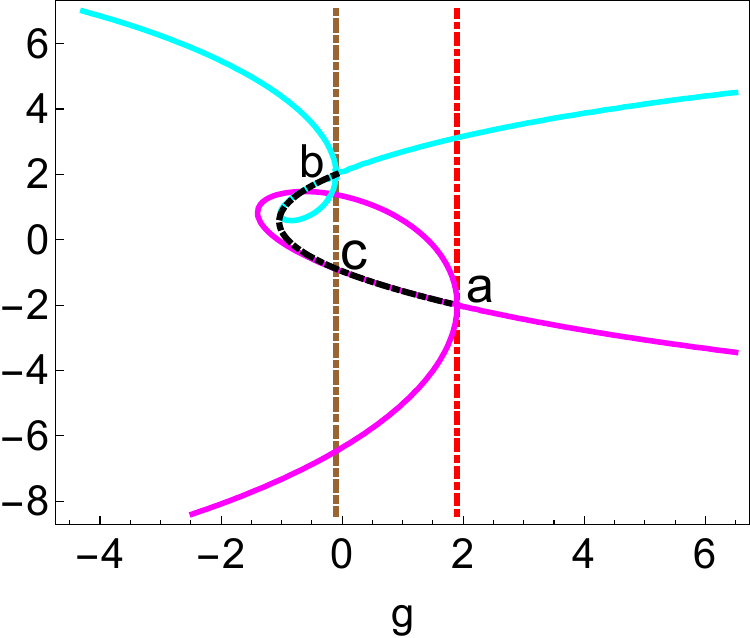}
\label{fig:overlap}}
\caption{ The flow diagrams for $k_0=0,\pi$: The background colour scheme, both in this figure and in Fig.\ref{fig-sshflows} 
shows the amplitude of the flow rate which decreases  from bright yellow (diverging)
 through light brown to its darkest shade which we shall call orange(vanishing).The blue lines represent flow lines for different initial conditions with arrows pointing in the direction of the flow. In all the flow diagrams, the flow rate diverges (vanishes) at the bright yellow (orange) regions. The vertical lines at $g=1.9$ (red) and $g=-0.1$ (brown) in the first two subfigures respectively  are the critical lines  which denote the 
topological phase transitions (compare with Fig.\ref{fig:pd}). The fixed point lines   where the flow rate vanishes are denoted by magenta and cyan dotted lines in the first two subfigures and as solid lines  in the third subfigure. The black dotted line is an array of gap closing points at non-HSPs  in Fig.\ref{fig:overlap}  and overlaps with part of the magenta and part of the cyan lines.  Only the unstable part of the fixed point lines overlaps with the black dotted critical line. The fixed point line in magenta intersects  the $k_0=0$ critical line (red) and also crosses itself at the same point  denoted by `a'. Similarly the fixed point  line in cyan for $k_0=\pi$ intersects the $k_0=\pi$ critical line (brown) and also crosses itself at the same point  `b'. Both of these are the multicritical points as shown in  Fig.\ref{fig:pd}. }
\label{fig:flows}
\end{figure*}

{\subsection{Flow equations, fixed lines and critical lines}}
 \label{sec:flow}


Using the RG procedure discussed in Sec.\ref{sec:curvature_rg} we now obtain the following flow  equations for the four coupling parameters  $\lambda$, $\lambda_2$, $g$ and $t$, around the high
symmetry points $k_0=0$ and $k_0=\pi$: 
\begin{eqnarray} 
\label{eq:flow}
\frac{d \lambda}{dl}  &=&   \frac{-1}{2(g-t\mp1)^2} {}  \alpha { (g, t, \lambda, \lambda_2)} \nonumber \\
\frac{d \lambda_2}{dl}&=& \frac{-\lambda_2}{2(g-t\mp1)(\lambda \pm 2 ) } {}  \alpha {(g, t, \lambda, \lambda_2)}\nonumber\\
\frac{dg}{dl} & = & \frac{1}{2(g-t\mp1)(\lambda \pm 2 ) } {}  \alpha {(g, t, \lambda, \lambda_2)} \nonumber \\
\frac{dt}{dl} & =&  \frac{-1}{2(g-t\mp1)(\lambda \pm 2 ) }{}  \alpha{(g, t, \lambda, \lambda_2)} 
\end{eqnarray}       
where 
\begin{align}
\label{Eq:fp}
\begin{split}
 \alpha(g, t, \lambda, \lambda_2) = {} & (\lambda \pm 8)(g-t\mp1)^2 + 2 \lambda_2^2 (\lambda \pm 2)^3 \\ {}& + 3(\lambda \pm 2) (g-t\mp1)(4t\pm1).
 \end{split}
 \end{align} 
  Here the upper sign is for $ k_0= 0 $ and lower sign for $ k_0=\pi$. From the above set of equations, it is obvious that $ dg/dl = - dt/dl $.  We are interested in flows of the parameters $(g, \lambda)$ under CRG. The other parameters in the model has been  kept fixed at $\lambda_2=0.85$ and $t=0.9$, since the qualitative features remain the same for other sets of values as well. 
  
  The flows of $(g, \lambda)$ as determined by CRG  Eq.\ref{eq:flow} are shown  in Fig.\ref{fig:flows}.  We find from the flow diagram  that CRG applied at HSP $k_0=0$ and $k_0=\pi$,  correctly captures all critical lines which are associated with gap closings at  both the HSPs.  The flow rate diverges on the vertical red line $g=t+1=1.9$ (brown line $g=t-1=-0.1$) which is a critical line where the  energy gap closes at $k_0=0$ ($k_0=\pi$).  As  is also obvious from Eq.\ref{eq:flow}, the RHS of all the RG equations diverge at $g=t+1$ for $k_0=0$ and at $g=t-1$ for $k_0=\pi$.  But there appears to be  no  direct way to discover the critical line on which energy gap closes at non-HSP $k$ points (the black dotted line in Fig.\ref{fig:pd} and Fig.\ref{fig:overlap}) from the  CRG method.

However a closer look at the flow diagram reveals that part of the fixed lines on which the flow rate vanishes  have  considerable overlap with the non-HSP gap closing critical line. The fixed lines can be obtained by setting the right side of flow Eq.\ref{eq:flow} to be zero and is given by $\alpha(g, t, \lambda, \lambda_2)=0$. The fixed line  for $k_0=0$ ($k_0=\pi$) is shown in  magenta (cyan) in Fig.\ref{fig:flowsk0} (Fig.\ref{fig:flowskp}). We note here that the region of overlap changes with the parameters
$\lambda_2$ and $t$ which we have kept fixed here, but the overlap never goes to zero. Only the unstable part of the fixed lines overlap with the critical line, which can be understood from the fact that both for the
critical line and for the unstable fixed line, we expect a small perturbation to go away from the line, as opposed to a stable fixed point, where  a perturbation brings it back. The fixed line for $k_0=0$ intersects 
the $k_0=0$ critical line and crosses itself at the same point which we denote by  `a'.  Similarly the fixed line for $k_0=\pi$ intersects the $k_0=\pi$ critical line and crosses itself at the same point which we denote
by  `b'. The two points  `a' and `b' are  precisely the multicritical points, which were earlier seen as points where three phases meet. Within the
curvature RG procedure, they are obtained as points where a critical line intersects an unstable fixed line. A direct evaluation of curvature function shows that $F(k_0, {\bf M})$ is indeterminate at these two points.  
In contrast, the multicritical point 'c' does not show the self-crossing
of the fixed point line. At this point the curvature function diverges at two non-HSP along with the $k=\pi$ point. \\

Thus, we see that the CRG procedure carried around the two HSPs is not only sufficient to find the critical lines associated with that HSP, but is also
able to identify the multi-critical point and also the  critical line associated with the non-HSPs, because the unstable fixed points overlap with it. 
In fact, by a suitable choice of parameters,  we can show that the almost the entire non-HSP critical line is reproduced by the unstable fixed point
lines at $k=0$ and/or $k=\pi$. The multi-critical point  `a' (`b') is  hence the  point where the band closing occurs at $k=0$ ($k=\pi$) and a non-HSP 

Here we have a series of gap closing non-HSP point in the BZ. These non-HSP k points are not isolated in the BZ.  One might think that a straight forward application of the CRG method at an arbitrary  non high symmetry point should be possible. 
However, $F({\bf k, M})$ is locally a maxima or minima  only at the HSPs  for any arbitrary choice of parameter values. 
This allows us to apply  Eq.\ref{eq:dr} at any point (except ${\bf M_c}$) in the parameter space without any obstruction. 
But at non-HSP's, the  curvature function  develops a peak only for a specific set of parameter values, corresponding to gap closing, where we can apply Eq.\ref{eq:dr}. Since, in general for the non-HSPs,  we do not have a 
priori knowledge of the critical
points in the parameter space, we are unable to apply the CRG method as a straightforward extension.


\vspace{0.2cm}

{\subsection{Correlation length and critical exponents}}
\label{kiteavexp}
%
As discussed in Sec.\ref{sec:curvature_rg}, the curvature function exhibits  scaling behavior near the topological transition given by $F(k_0,{\bf M})\sim |{\bf M}-{\bf M}_c|^{-\gamma}$. From the Lorentzian form in Eq.\ref{Eqn:oz}, the width of the curvature function at $k_0$ goes to zero in this limit, resulting in a scaling form $\xi_{k_0} \sim |{\bf M}-{\bf M}_c|^{-\nu}$. The exponents $\nu$ and $\gamma$ are the critical exponents characterizing the TPT.  
 
  The exponents can be extracted by evaluating the curvature function and its Lorentzian expansion close to the high symmetry points $k_0=0$ and $\pi$. The curvature function has the form   $ F(k_0, {\bf M})=\pm  \lambda_2 (\lambda \pm 2)/(g - t \mp 1)$, where the upper (lower)  sign is for $k_0=0$ ($k_0=\pi$). Near the critical points, i.e. $g\rightarrow t\pm 1$, $F(k_0, {\bf M})$ diverges as
  \begin{eqnarray}
   F(k_0, {\bf M})\sim  \frac{1}{|g - t \mp 1|}, 
 \label{eq:lengthf}
    \end{eqnarray}
    giving $\gamma=1$.  Further, the Lorentzian form of the   curvature function around $k_0=0$ and $\pi$ yields
  \begin{eqnarray}
  \xi_{k_0}({\bf M})= \Big{ |} \frac{\alpha(g, t, \lambda, \lambda_2) }{2(g-t \mp 1)^2 (\lambda  \pm 2) }   \Big{|}^{1/2} \sim  \frac{1}{| {g - t \mp 1} |} \qquad  \label{eq:length}
  \end{eqnarray}
giving $\nu=1$.

\vspace{0.5cm}

\section{CRG analysis of the extended Su-Schrieffer-Heeger chain}
\label{sec:ssh}
\subsection{The model and its phase diagram}
\label{sec:model-ssh}
The second model we consider is  the well-studied Su-Schrieffer-Heeger (SSH) model\cite{}, but now with the introduction of third-nearest-neighbour (NNNN) hopping terms  $T_1$  and  $T_2$\cite{Rufo2019}, with the Hamiltonian given by
\begin{align}
H & = t_1\sum_i (c^{\dagger}_{A,i} c_{B,i} + h.c) +  t_2 \sum_i (c^{\dagger}_{B,i} c_{A,i+1} + h.c)  \nonumber \\
  & + T_1\sum_i (c^{\dagger}_{A,i} c_{B,i+1} + h.c) + T_2  \sum_i (c^{\dagger}_{B,i} c_{B,i+2} + h.c) .
\end{align}
The presence of the third  nearest neighbour allows for a richer phase diagram with multi-critical points as shown by  Ref.\onlinecite{Rufo2019}.
The Hamiltonian can be readily diagonalised by Fourier transformation, and the Hamiltonian turns out to be the same as that in Eq.\ref{2stateham} with 
\begin{align}
  d_1(k) = & t_1 + t_2 \cos k + T_1\cos k + T_2 \cos 2k \nonumber \\
  d_2(k) = & t_2 \sin k - T_1 \sin k + T_2 \sin 2k  \nonumber \\
  d_3(k) = & 0.
\end{align}
Here, the Pauli matrices however represent the sub lattice degrees of freedom (instead of particles and holes in the Kitaev chain).
This 1D model belongs to the symmetry class  AIII,  with chiral symmetry  $S=\sigma_3$ and  time reversal symmetry $T$  (this  being 
the analog of complex conjugation in the Kitaev chain). Here again, the model exhibits phases that are distinguished by the winding 
number $W$ given by 
\begin{align}
W  =  \frac{1}{2\pi} \int dk \hspace{0.1cm} F(k, {\bf M}),  
\end{align}
with  the curvature function being  given by 
 \begin{align}
  F(k, {\bf M}) = & \frac{1}{2i} Tr(\sigma_3 H_k^{-1} \partial_k H_k) = \frac{d_1 \partial_k d_2 - d_2 \partial_k d_1}{E^2(k)}. 
  \end{align}
The eigenvalues of $H_k$ are $E(k) = \pm \sqrt{d_1^2  +  d_2^2} $  and the parameters are ${\bf M} = (t_1, t_2, T_1, T_2)$. Without the third nearest neighbour, the model is well-known\cite{}  to have has topologically distinct phases with winding number $W=1$ and $W=0$,  
 when  $|t_2|>|t_1|$  and  $|t_2|<|t_1|$ respectively.\\

\begin{figure*}[htb]
\captionsetup{position=bottom}
\subfloat[ ]{\includegraphics[width=0.28\linewidth, height=0.28\linewidth]{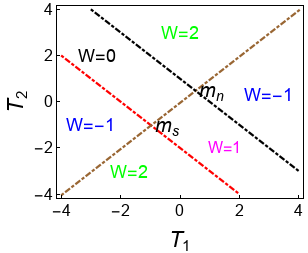}}
\subfloat[ $k_0=0$]{\includegraphics[width=0.35\linewidth, height=0.28\linewidth]{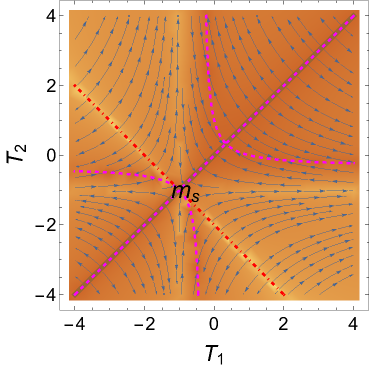}}
\subfloat[$k_0=\pi$ ]{\includegraphics[width=0.35\linewidth, height=0.28\linewidth]{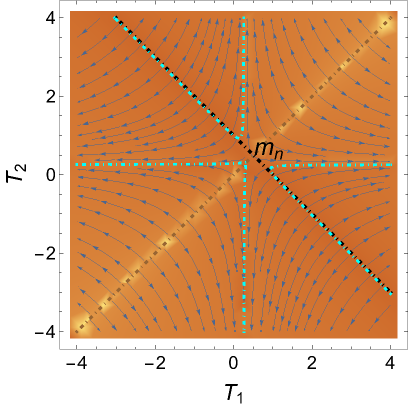}}\\
\subfloat[ ]{\includegraphics[width=0.28\linewidth, height=0.28\linewidth]{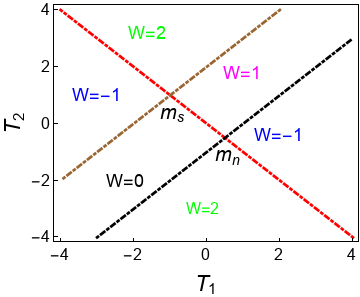}}
\subfloat[$k_0=0$ ]{\includegraphics[width=0.35\linewidth, height=0.28\linewidth]{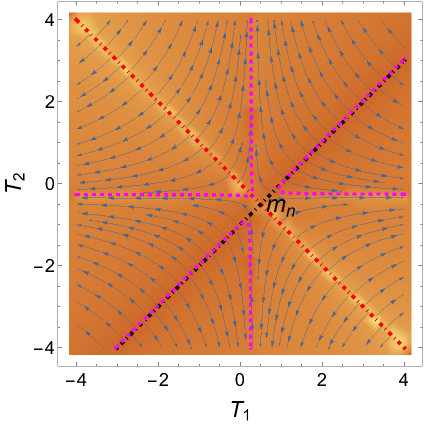}}
\subfloat[ $k_0=\pi$]{\includegraphics[width=0.35\linewidth, height=0.28\linewidth]{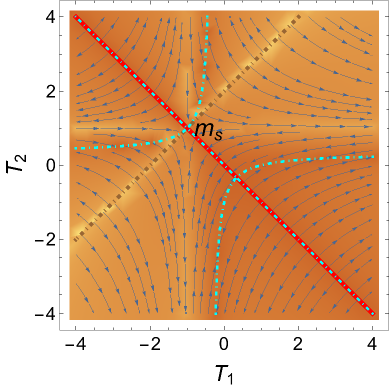}} 
\caption{The phase diagram and RG flows for the extended SSH model around the HSPs $k_0=0$ and $k_0=\pi$.  (i) The panels in the top row are 
for   $t_1=t_2=1$ and (ii)  the panels in the bottom row are  for $t_2=-t_1=1$. (a) and (d) are the phase diagrams obtained from an exact diagonalisation.
(b) and (e) are the flow diagrams obtained around the $k_0=0$ HSP and (c) and (f) are the flow diagrams around the $k_0=\pi$ HSP. 
In all the diagrams, the red dotted line is the critical line on which  gap closes at $k=0$, the brown dotted line is due to gap closing at $k_0=\pi$ and the black dotted line is associated with   gap closing at the  non-HSP $k=2\pi/3$ ($k=\pi/3$) for top (bottom) panel.
In all the flow diagrams,  the flow rate diverges (vanishes) on the bright yellow regions (orange ) which correspond the critical lines (fixed lines). The fixed point  lines where the flows stop are shown in magenta (cyan)  for $k_0=0$ ($k_0=\pi$). Note that in all the flow diagrams there is overlap of the critical lines and part of the fixed point lines and also that they cross
another  critical line, denoting the existence of multi-critical points.
In (b), the bright yellow region overlaps with the red dotted line as expected. But there is also a part  of the magenta line which corresponds the unstable fixed lines (the forward  leaning diagonal part), which completely overlaps with the brown dotted critical line, and we get the multi-critical point $m_s$.  In (c), the bright yellow region overlaps with the brown dotted line as expected. But there is also an almost complete overlap of part of the cyan line which  corresponds  the unstable fixed line(the backward leaning diagonal  part), with the black dotted line.  In (e), once again, the bright yellow region overlaps with the red dotted line. But here, there is an almost complete overlap of the magenta line (forward leaning diagonal) with the black dotted line and in (f), besides
the overlap of the bright yellow region  with the brown dotted line, we have a complete overlap of the cyan line (backward leaning diagonal) with the red dotted line and we get the multi-critical point $m_s$. In all the cases, the  unstable portion of the  fixed line overlaps with critical line. }
\label{fig-sshflows}
\end{figure*}

As  shown in Ref.\onlinecite{Rufo2019}, adding NNNN hopping terms $T_1$  and $T_2$ allows for phase diagrams with multi-critical points.
We have obtained the phase diagram  as shown in   Fig.\ref{fig-sshflows}(a) for  fixed $t_1=t_2=1$. We have also obtained the phase diagram for 
$t_2=-t_1$. Similar to the Kitaev chain, it can be shown that the model not only has critical lines coming from the gap closing at the HSPs $k_0=(0,\pi)$, but also from a gap closing at a non-HSP $k_0=2\pi/3$ for $t_1=t_2$ 
($k_0=\pi/3$ for $t_2=-t_1$). In parameter space, there are the   three critical lines  $T_1 + T_2 + t_1 + t_2 = 0$ (gap closes at $k=0$),  $-T_1 + T_2 + t_1 - t_2 = 0$ (gap closes at $k=\pi$) and  $T_1 + T_2 + t_2 - 2t_1 = 0$ (gap closes at $k=2\pi/3$) 
for $t_1=t_2$. For $t_2=-t_1$, the only difference is that the third critical line is given by   $T_1-T_2+t_2+2t_1=0$ (gap closes at $k=\pi/3$). Note that the non-HSP critical lines are parameter dependent and are different for $t_1=t_2$ and $t_1=-t_2$.


{\subsection{Flow equations, fixed lines and critical lines}}

Here again, we apply the RG procedure discussed in  Sec.~II and obtain 
the following flow equations at  the two HSPs $k_0=0$ and $k_0=\pi$,  for the four parameters $t_1,t_2,T_1,T_2$.

\begin{align}
\label{eq-sshflowt1} &    \frac{dt_1}{dl} = \frac{\alpha(t_1, t_2, T_1, T_2)}{2(t_1 \pm t_2 \pm T_1 + T_2)(\pm t_2 \mp T_1 + 2T_2)}  \\
 \label{eq-sshflowt2} &   \frac{dt_2}{dl} = \frac{-\alpha(t_1, t_2, T_1, T_2)}{2(t_1 \pm t_2 \pm T_1 + T_2)(\pm t_1 + 2T_1 \mp T_2)} \\ 
 \label{eq-sshflow1} &    \frac{dT_1}{dl} = \frac{\pm \alpha(t_1, t_2, T_1, T_2)}{2(t_1 \pm t_2 \pm T_1 + T_2)(t_1 \pm 2t_2 + 3T_2)}  \\
  \label{eq-sshflow2} &   \frac{dT_2}{dl} = \frac{-\alpha(t_1, t_2, T_1, T_2)}{2(t_1 \pm t_2 \pm T_1 + T_2)(2t_1 \pm t_2 \pm 3T_1)}, 
\end{align}
with  $ \alpha(t_1, t_2, T_1, T_2)$ being  given by
\begin{align}\label{eq-sshfixed}
& \alpha(t_1, t_2, T_1, T_2)  = - (\mp t_2 \pm T_1 - 8T_2)(t_1 \pm t_2 \pm T_1 + T_2)^2  \nonumber \\
&     +  3(\pm t_2 \mp T_1 + 2T_2)(\mp t_2 \mp T_1 - 4T_2)(t_1 \pm t_2 \pm T_1 + T_2) \nonumber \\
&  +  2(\pm t_2 \mp T_1 + 2T_2)^3.
\end{align} 
The upper(lower) sign is  for $k_0=0$ $(\pi)$.
For convenience, to show the RG flows,  we  choose to set  $(t_1, t_2)$  as fixed parameters and take $(T_1, T_2)$ as  the tuning parameters. 
The flows of $T_1$ and $T_2$ as governed by the equations Eq.\ref{eq-sshflow1} and Eq.\ref{eq-sshflow2}, are shown in  Fig.\ref{fig-sshflows}. 

Just as before, here again we note that the flow diagram correctly captures all the critical lines on which the gap closes at the HSPs $k_0=0$ and
$k_0=\pi$ for two sets of the fixed parameters $t_1=t_2=1$ and $t_2=-t_1=1$. These are the lines where the flow rate diverges. But a  closer look at  
Fig.\ref{fig-sshflows}  reveals that  the CRG applied at a particular HSP $k_0$ is also able to capture a large fraction of the other critical lines which are associated with  the gap closing at a different $k_0$ point,  (which could either be another HSP or a non-HSP), as unstable fixed point lines.

Let us look carefully at the panels in the top row of Fig.\ref{fig-sshflows}. As explained in the caption, the left panel shows the critical lines
for the two HSPs, $k_0=0,\pi$ and also for the non-HSP $k_0=2\pi/3$. The middle panel shows the RG flows around
the fixed point at $k_0=0$. Here, we can clearly see that it not only reproduces the $k_0=0$ critical line (where the flow equations diverge),
 but the line passing through the  unstable fixed points (where the flow equation vanishes) also reproduces the line where the gap closes
 at the HSP, $k_0=\pi$. Further, the two critical lines cross at the multi-critical point ${\bf m}_s$, which is thus a point where the gap closes
 simultaneously at $k_0=0$ and $k_0=\pi$ denoting two different phase transitions. A direct evaluation shows that curvature function is indeterminate at $k_0=0$  
 when ${\bf M}={\bf m}_s$.  Similarly, the panel on the right which denotes the flow equations around the HSP $k_0=\pi$ clearly
 reproduces the $k_0=\pi$ critical line (where the flow equations diverge), but also the unstable fixed point line reproduces the $k_0=2\pi/3$ 
 critical line almost completely, except for a small portion near the crossing point, which is the multi-critical point ${\bf m}_n$. We have in fact, checked
 that the overlap becomes more and more complete, when the parameters $t_1=t_2 \to 0$. Note that in this limit, the red and black dotted critical lines in 
 Fig.5a merge into one, and  the model reduces to a SSH model  (with $T_1$ and $T_2$).  
 Here, the multi-critical point ${\bf m}_n$  represents gap closings
 at both $k_0=\pi$ and $k_0=2\pi/3$ and the curvature function is indeterminate at $k_0=2\pi/3$.
  
  In the lower panels, the values of $t_1$ and $t_2$ have been changed and it is now plotted for $t_2=-t_1=1$.
 Here, again, we see that the middle panel, with RG flows at $k_0=0$ not only reproduces the critical line, but the unstable fixed point line 
 reproduces the $k_0=\pi/3$ critical line to a great extent, and which can be made almost complete by choosing the parameters $t_2=-t_1 \to 0$,
 (where again, the  model reduces to a different SSH model); 
 and on the right panel, besides reproducing the $k_0=\pi$ critical line, the unstable fixed point line
 also reproduces the $k_0=0$ critical line.
 
 In principle, in this model, it would be possible for us to apply the CRG procedure here to the isolated non-HSP $k_0=2\pi/3$ (for $t_1=t_2$)
 and for the other non-HSP $k_0=\pi/3$  (for $t_1=-t_2$) as well,  since unlike the earlier model, here we know the critical point in parameter space.
 However, as we can see, the non-HSP is parameter dependent, and we do not expect to get any further information from those
 flow equations that we have not already obtained. So we do not do that here.
The essential point that we wish to emphasize here that the RG flows around any particular critical point also
  reproduces the phase boundaries around other critical points. In other words, the RG equations around a point $k_0$, which
  requires only the knowledge of the curvature function in its neighbourhood, is sufficient to give us the phase boundaries in the entire Brillouin zone.

\vspace{0.2cm}
\subsection{Correlation length and critical exponents }
The scaling behaviour of the curvature function  can be obtained by evaluating it around the two HSPs where the curvature function  diverges. We get the following  forms for $F(k_0, {\bf M}) = - (\mp T_1 + 2T_2 \pm t_2)/(\pm T_1 + T_2 + t_1 \pm t_2)$ where the upper (lower) sign is for for $k_0=0$ ($k_0=\pi$). Close to  criticality,    $\pm T_1 + T_2 \to - t_1 \mp t_2$, the curvature function diverges as 
\begin{align}
F(k_0, {\bf M}) \sim \frac{1}{|\pm T_1 + T_2 + t_1 \pm t_2 |} , 
\end{align}
which gives the critical exponent $\gamma=1$. As for the earlier case,  we can compute the length scale $\xi_{k_0}$  by  bringing $F(k_0 + \delta k, {\bf M})$  in the Lorentzian form as  given in Eq.\ref{Eqn:oz}  and we get the following expression - 
\begin{align}
\xi_{k_0}({\bf M}) =  {\Big{|} \frac{\alpha(t_1, t_2, T_1, T_2)}{(t_1 \pm t_2 \pm T_1 + T_2)^2 (\pm t_2 \mp T_1 + 2T_2)} \Big{|}}^{1/2}.
\end{align} 
Close to criticality,   $i.e.$ when  $\pm T_1 + T_2 \to - t_1 \mp t_2$,  $\xi_{k_0}$ has the same divergence structure as the curvature $F(k_0, {\bf M})$:   $\xi_{k_0}({\bf M}) \sim {|\pm T_1 + T_2 + t_1 \pm t_2 |}^{-1}$  which makes the critical exponent $\nu = 1$ as well.

\vspace{0.2cm}
\section{The parameter landscape}

In the earlier sections, we explicitly studied two models in one dimension both of which had multicritical points where multiple topological
phases met. In both models, we applied the CRG procedure around the HSPs and studied the RG flows and the fixed point lines.
Here, we present the conclusions which could possibly be generalised to other models. 

In the general parameter space, the models not only have gap closings at the HSPs, but also gap closings and phase transitions at non-HSPs.
We found that just  by studying the CRG around the HSPs and analysing their fixed point structure, we could find out possible multicritical
points in the models and also the possible  topological phase transitions that occur at non-HSPs.  Essentially, when we  looked at the CRG procedure around a particular critical point, we found that other critical points appear as unstable fixed points.  We can understand this as follows.
In the parameter landscape,  at the point of phase transition ${\bf M} = {\bf M}_c$, the bulk band closes somewhere in the Brillouin zone resulting
in a divergent curvature function.  Thus, if there are many critical points, the curvature function has many divergences at different points $k_0$ in the BZ 
and for different parameters ${\bf M_c}$. Thus in this multi-dimensional  ( including $k$ as a label) space, there are many divergences and it
is clear that if we could look at all possible flow diagrams, then the CRG procedure from any one of the critical points would lead you to the
others as possible unstable fixed points.    Interestingly, even when we look at a restricted parameter space (for instance, in both the
models, we vary only two parameters at a time), we are able to infer the existence of all the other critical points in the theory.
In this landscape picture, the multicritical point is essentially a point where for a given set of parameters, there are two different gap closings
allowing for two different phase transitions. 

 \section{Discussion and conclusions}
 \label{sec:dc}

 In this paper, we have studied the multi-critical points of topological phase transitions using the CRG procedure. We have clarified the question
 of the apparent meeting of fixed line and critical line at the multi-critical point, essentially by showing that for a wide range of
 parameters, the unstable fixed point lines of the CRG equations  of a particular gap closing critical point actually overlaps with the 
 critical lines of other gap closing critical points in the Brillouin zone. Interestingly, by choosing appropriate parameters, we are able
 to increase the overlap to be almost complete.
 
  As we have explained earlier, if we could study the flows in the full multi-parameter space and if 
 the CRG equations could be extended to higher order, we could expect that some of the overlaps between the unstable fixed
 point lines  of a given  critical gap closing momentum and the critical lines at other possible gap closing momenta, which we have
 obtained as incomplete, would probably be complete.
 
 We have studied two models in detail - the extended Kitaev model and the extended Su-Schrieffer-Heeger model.
 While these models are in one-dimension and are exactly solvable, a similar analysis should be possible for interacting models
 and models in higher dimensions as well.  Since the CRG procedure around a HSP yields information about other phase transitions
 occuring due to gap closures at other momentum points and even multi-critical points, we expect that this would be a useful
 technique to apply when exact computations are not possible. However, one caveat to keep in mind is that although
 if we could study the flows in the full multi-parameter space and  the CRG equations could be extended to higher order, we
 would expect to get all the phases, the fact that we are able to get so much information from a restricted parameter space
 may be only true for simple models in one dimension.
 
 Going towards the future, we expect  further generalisations to models  in different dimensions and in different symmetry classes.
 A recent review\cite{Chitra2019} tries to give a unified picture of topological phase transitions in a variety of static and periodically driven systems, and in  both weakly and strongly interacting systems with the aim of classifying these transitions using standard concepts
 of critical exponents and universality classes. It would be of interest to see whether a similar analysis of  multicriticality in  all those models is possible.

 \acknowledgements {We would like to thank Paolo Molignini for useful correspondence. We would also like to acknowledge the workshop `Novel Phases of quantum Matter'  at ICTS for hospitality during the course of this work.}

\appendix


 \section{Curvature function in the whole BZ for the Extended Kitaev model}

 \renewcommand{\thefigure}{A\arabic{figure}}
 \setcounter{figure}{0}
 
\begin{figure}[h]
 \captionsetup{position=bottom}
\subfloat[ ]{\includegraphics[width=0.49\linewidth, height=0.35\linewidth]{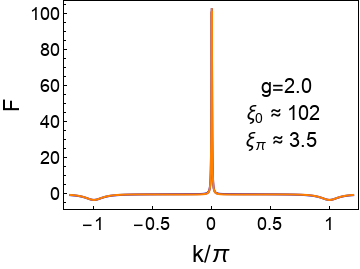}
\label{fig.lz1_8}}
\subfloat[ ]{\includegraphics[width=0.49\linewidth, height=0.35\linewidth]{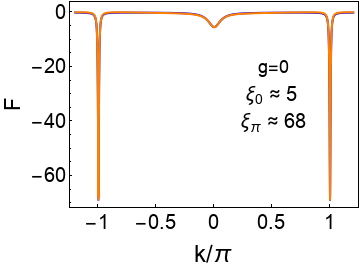}
\label{fig.lz0}}\\
\subfloat[ ]{\includegraphics[width=0.49\linewidth, height=0.35\linewidth]{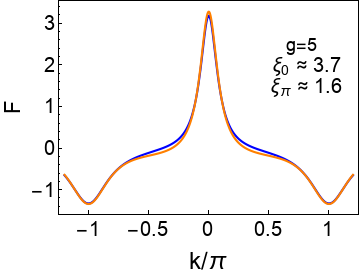}
\label{fig.lz5}}
\caption{The orange line denotes the exact curvature function  and  the blue line denotes the  sum of the Lorentzians at two HSPs $k_0=0, \pi$. In subfigures (a) and (b), where the parameter values are close to criticality, the blue and orange lines overlap everywhere. In all figures we take $\lambda=10$. The length scales $\xi_0$ and $\xi_{\pi}$ are computed from the Lorentzians. }
\label{fig.lz}
\end{figure}

 Here, we study the curvature function as well as the sum of the Lorentzians at the HSPs $k_0=0, \pi$ in Eq.\ref{Eqn:oz}  in the entire Brillouin zone  for a few representative  parameter regimes.
Note that the sum of the Lorentzians is not just a very good fit to the curvature function close to criticality (e.g. in Figs.\ref{fig.lz}(a) and
(b)), but also for parameter values which are not close to either the $k_0=0$ or $k_0=\pi$ critical values Fig.\ref{fig.lz}(c).

\section{Curvature function in the whole BZ for the extended Su-Schrieffer-Heeger model}
\begin{figure}[h]
 \captionsetup{position=bottom}
\subfloat[ ]{\includegraphics[width=0.48\linewidth, height=0.35\linewidth]{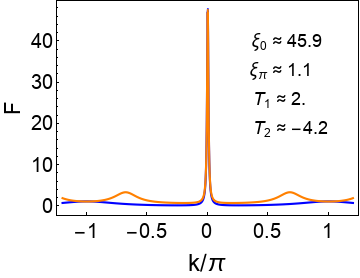}
\label{fig.ssh-k=0}}
\subfloat[ ]{\includegraphics[width=0.48\linewidth, height=0.35\linewidth]{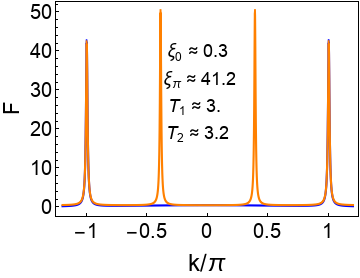}
\label{fig.ssh-k=pi}}\\
\subfloat[ ]{\includegraphics[width=0.48\linewidth, height=0.35\linewidth]{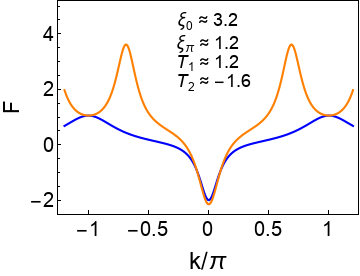}
\label{fig.ssh-k=n}}
\caption{The orange line denotes the exact curvature function  and  the blue line denotes the  sum of the Lorentzians at both HSPs $k_0=0$ and $k_0=\pi$. 
In subfigures (a) where the parameter values are close to $k_0=0$ critical points, the blue and orange lines overlap everywhere except at $k=\pm 2\pi/3$. In In subfigures (b), where parameters are close $k_0=\pi$ critical line,  the exact curvature function shows additional divergent peak  at other k point in addition to the HSP $k_0=\pi$. In figure (c), where parameters are not close to any of the critical lines, the Lorentzian fit of two HSPs  is far from matching the exact curvature function at  $k=\pm 2\pi/3$.
The length scales $\xi_0$ and $\xi_{\pi}$ are computed from the Lorentzians. }
\label{fig.sshlz}
\end{figure}

Here, we study the   curvature function for the extended SSH chain, which is shown in  Fig.\ref{fig.sshlz}  along with the Lorenztian fits 
at  both the HSPs $k_0=0$ and $k_0=\pi$ for a few representative
parameter values. In this model,  we also have an isolated gap closing point at a  non-HSP  $k=\pm 2\pi/3$ in the BZ (for $t_1=t_2=1$). As  is clear  from the figures,  just taking the Lorentzian fits around the HSPs is not sufficient to obtain
the curvature function everywhere in the BZ. 

\bibliographystyle{apsrev}


\end{document}